# GENERAL PROPERTIES OF OPTICAL HARMONIC GENERATION FROM A SIMPLE OSCILLATOR MODEL

F. Bassani and V. Lucarini
Scuola Normale Superiore, 56100 Pisa

## ABSTRACT

The nonlinear oscillator model allows a basic understanding of all nonlinear processes and can be adopted to analyse optical vibrational modes and electronic transitions in molecules and crystals, in order to derive general properties of harmonic generation to all orders. In particular, we obtain Kramers-Krönig relations and sum rules referred to all momenta of the susceptibility, and Miller's empirical rules to all orders. Since the above properties only depend on time causality and not on the specific microscopic model, they can also be adopted for the quantum mechanical description, substituting in the classical expressions the derivatives of the potential with their expectation values.

## II. INTRODUCTION

Since the advent of laser, nonlinear optical effects have been extensively studied both theoretically and experimentally [1]. The possibility of producing particularly high electric fields of very short duration has recently increased the interest in harmonic generation processes [2]. Since the first detection of second harmonic generation [3], a continuous progress along similar lines has allowed the detailed investigation of higher harmonics [4]. In the case of atoms, using high power sources, odd order harmonics as high as the 135$^{th}$ have been obtained [5].

Kramers-Krönig (K.K.) relations and related sum rules have been proved for second harmonics [6] and third harmonics [7], using the Kubo expression [8] for the response function. The above sum rules have allowed a derivation of Miller's empirical rules for second harmonic generation and the calculation of Miller's constant [9], using a simplified quantum mechanical treatment. Sum rules can be in general related to the ground state properties only [10].

The purpose of the present contribution is to obtain similar results for harmonics of any order by using the classical anharmonic oscillator model in the presence of one external electric field of a given frequency. Since the basic ingredients are the analytical properties of the susceptibility $\chi^{(n)}(\omega...\omega)$ and its asymptotic behavior, the results obtained in this way can be interpreted from a more general point of view.

In section II we describe our model and give (the prescriptions by obtaining) analytical solutions to all orders.

In section III we use the analytical properties and the asymptotic behavior of the nonlinear susceptibility to derive appropriate Kramers-Krönig (K.K.) relations and sum rules for the momenta of the susceptibility up to the highest momentum.

In section IV we show how Miller's rules can be obtained for the harmonic generation to all orders and how they are interconnected.

In section V we discuss the application of our general results to ionic and electronic motion.

In section VI we present our conclusions.

## II. ANHARMONIC OSCILLATOR MODEL DRIVEN BY A FIELD OF GIVEN FREQUENCY

We follow the idea of the Lorentz oscillator and consider a general potential energy $mV(x)$ expanded around the equilibrium position of moving charges.

Let the Hamiltonian of the forced oscillator, for simplicity taken in one dimension, be:

$$H = \frac{p^2}{2m} + \sum_{n=2}^{\infty} \frac{m}{n!} \left[ \frac{\partial^n V(x)}{\partial x^n} \right]_0 x^n - eE_\omega x \, , \tag{1}$$

where

$$E_\omega = E_0 e^{-i\omega t} + c.c. \, , \tag{2}$$

where we have considered only one beam of well defined frequency because we confine ourselves to harmonic generation and we have neglected spatial dispersion (dipolar approximation).

When we include damping, the equation of motion becomes:

$$\ddot{x} + \gamma \dot{x} + \omega_0^2 x + \sum_{n=3}^{\infty} \left[ \frac{\partial^n V(x)}{\partial x^n} \right]_0 \frac{x^{n-1}}{(n-1)!} = \frac{eE_\omega}{m} \, . \tag{3}$$

The solutions can be obtained by successive iterations, each iterative process giving a higher harmonic.

The procedure requires expressing the solution as a sum of terms of decreasing magnitude:

$$x = x_{(1)} + x_{(2)} + x_{(3)} + \ldots \, , \tag{4}$$

each term being responsible for various effects, among which the production of the corresponding harmonic. Substituting (4) in (3) and separating the various orders terms we obtain the solutions for harmonic generation.

The linear contribution is:

$$x_{(1)} = \frac{e}{m} \frac{1}{\omega_0^2 - \omega^2 - i\gamma\omega} E_\omega \, , \tag{5}$$

which gives the usual first order susceptibility:

$$\chi^{(1)}(\omega) = \frac{e^2 N/m}{D(\omega)} \, , \tag{6a}$$

with the definition:

$$D(\omega) = \left( \omega_0^2 - \omega^2 - i\gamma\omega \right) \, . \tag{6b}$$

The second order susceptibility for second harmonic, as also given by many authors [1], is:

$$\chi^{(2)}(\omega,\omega) = -\frac{1}{2!}\left[\frac{\partial^3 V(x)}{\partial x^3}\right]_0 \frac{e^3 N/m^2}{D^2(\omega)D(2\omega)} \ . \tag{7}$$

The third order susceptibility for third harmonics is:

$$\chi^{(3)}(\omega,\omega,\omega) = -\frac{1}{3!}\left[\frac{\partial^4 V(x)}{\partial x^4}\right]_0 \frac{e^4 N/m^3}{D(3\omega)D^3(\omega)} + \left[\frac{\partial^3 V(x)}{\partial x^3}\right]_0^2 \frac{e^4 N/m^3}{D(3\omega)D(2\omega)D^3(\omega)} \ . \tag{8}$$

We can observe that the first term has a slower decreasing behavior at infinity ($\propto \omega^{-8}$), while the second term goes to zero much faster ($\propto \omega^{-10}$). In previous analyses [1],[11] the second term had not been considered. The higher order susceptibilities for harmonic generation can also be obtained. They are:

$$\chi^{(n)}(\omega,\ldots,\omega) = -\frac{1}{n!}\left[\frac{\partial^{n+1} V(x)}{\partial x^{n+1}}\right]_0 \frac{e^{n+1} N/m^n}{D(n\omega)D^n(\omega)} - \frac{1}{D(n\omega)} \times$$

$$\times \sum_{i,k,l} \left( \frac{1}{e^{\sum_{u=1}^{l} k_u - 1} N^{\sum_{p=1}^{l} k_p - 1}} \left[\frac{\partial^{\left(\sum_{y=1}^{l} k_y + 1\right)} V(x)}{\partial x^{\left(\sum_{y=1}^{l} k_y + 1\right)}}\right]_0 \prod_{j=1}^{l} \frac{\left(\chi^{(i_j)}(\omega\ldots\omega)\right)^{k_j}}{k_j!} \right) , \tag{9a}$$

with the constraints:

$$\sum_{j=1}^{l} k_j i_j = n , \tag{9b}$$

and:

$$\begin{array}{l} \forall s,s'|1 \leq s < s' \leq l \\ (1 \leq i_s < i_{s'} < n) \wedge (0 \leq k_s < n) \end{array} ; \tag{9c}$$

note that the second one in (9c) is needed to separate the first term in (9a) from teh rest of the terms in the summation.

We can observe that in (9a) only the first term is relevant for the asymptotic behavior, because all the other contributions go to zero much faster at infinity. Furthermore, in case of inversion symmetry, only odd order terms of the susceptibilities are different from zero because in the even order susceptibilities the first term vanishes and the terms in the summation also vanish since they always contain as a factor at least one odd order derivative of the potential which obviously vanishes.

### III. KRAMERS-KRÖNIG RELATIONS AND SUM RULES

We have proved in all generality that the susceptibilities $\chi^{(n)}(\omega,\omega..\omega)$ are analytic functions in the upper complex $\omega$ plane, due to time-causality of the general response function [12]. This can be seen to hold obviously for all preceding expressions, the reason being that time causality has been essentially introduced in the equation of motion.

As a consequence a number of K.K. relations hold for each harmonic generation process when the different asymptotic behaviour is considered. All the K.K. relations for each process can thus be obtained and are the following:

$$\begin{cases} \omega^{2\alpha} \operatorname{Re} \chi^{(n)}(\omega..\omega) = \dfrac{2}{\pi} \int_0^\infty \dfrac{\omega'^{2\alpha+1} \operatorname{Im} \chi^{(n)}(\omega'..\omega')}{\omega'^2 - \omega^2} d\omega' \\ \omega^{2\alpha-1} \operatorname{Im} \chi^{(n)}(\omega..\omega) = -\dfrac{2}{\pi} \int_0^\infty \dfrac{\omega'^{2\alpha} \operatorname{Re} \chi^{(n)}(\omega'..\omega')}{\omega'^2 - \omega^2} d\omega' \end{cases}$$

$$\text{with } 0 \leq \alpha \leq n \qquad . \qquad (10)$$

The number of K.K. relations for each case depends on the order of harmonic because the exponent $2\alpha$ has to be such that $\omega^{2\alpha} \chi^{(n)}(\omega..\omega) \in L^2(-\infty,\infty)$.

With respect to the linear case, where K.K. relations have been systematically exploited [13], the K.K. relations (10) have not been extensively applied. We can notice that, of the many relations obtained, those with lower $\alpha$ values have a very rapid convergence while the highest $\alpha$ values require the knowledge of the entire spectrum. For this reason so far only the $\alpha = 0$ relations have been used. In the case of third harmonics Hasegawa *et al.* [14] have shown that the phase of $\chi^{(3)}(\omega,\omega,\omega)$ can be extracted by a measure of its modulus over a small frequency interval by using the $\alpha = 0$ K.K. relations. The other dispersion relations can be of value for analysing and extrapolating all data on high harmonic generation.

With the above dispersion relations (10) and the asymptotic behavior given by (9), sum rules can be obtained by setting $\omega \to 0$ in expressions (10) or by considering their asymptotic behavior as obtained from the superconvergence theorem [12], [13]. When such asymptotic behavior is made to agree with that obtained from expression (9) we obtain the following sum rules:

$$\int_0^\infty \omega'^{2\alpha} \operatorname{Re} \chi^{(n)}(\omega'...\omega')d\omega' = 0 \qquad \text{with} \ \ 0 \leq \alpha \leq n$$

$$\int_0^\infty \omega'^{2\alpha+1} \operatorname{Im} \chi^{(n)}(\omega'...\omega')d\omega' = 0 \qquad \text{with} \ \ 0 \leq \alpha \leq n-1$$

$$\int_0^\infty \omega'^{2\alpha+1} \operatorname{Im} \chi^{(n)}(\omega'...\omega')d\omega' = \frac{\pi}{2}\frac{1}{n!}\frac{e^{n+1}N}{m^n}\left[\frac{\partial^{n+1}V(x)}{\partial x^{n+1}}\right]_0 \frac{(-1)^{n+1}}{n^2}$$

$$\text{with} \ \alpha = n \ . \qquad (11)$$

Higher momenta of the susceptibility than the ones considered above do not give convergence.

As proved in detail for the case of second harmonics and of third harmonics, the validity of the above described dispersion relations and sum rules extends beyond the simple classical model adopted, here provided the external potential V(x) of the anharmonic oscillator is substituted with the effective crystal potential for the electrons. It has also been proved [6] [7] that, in the case of the quantum mechanical treatment, instead of the derivatives at the equilibrium position their expectation values on the ground state have to be considered.

When we also consider the possibility of different directions of the applied fields and of the polarization, we obtain the usual tensorial relations in the sum rules and in the K.K. relations, the type of tensorial matrices being determined in the usual way by the symmetry of the potential.

The above sum rules as the K.K. relations (10) provide a large number of constraints which must be obeyed by any theoretical model of harmonic generation processes. Since in most materials the theoretical ab initio calculations of the susceptibilities are hardly possible, such constraints may be used to determine the relevant parameters inside a particular quantum model, as in the case of a single resonance frequency for $\chi^{(2)}(\omega,\omega)$ [6], [7].

The nonlinear sum rules for each specific case have been used in experimental studies of the optical nonlinear response of atomic cesium vapor [15].

## IV. DISCUSSION OF MILLER'S RULES AND FURTHER APPLICATIONS

Miller proposed in 1964 an empirical rule connecting the second harmonic susceptibility $\chi^{(2)}(\omega,\omega)$ with the first order susceptibilities $\chi^{(1)}(\omega)$ [16]. Subsequently Iha and Bloembergen [17] have extended Miller's rule to the case of third harmonic generation. Miller's constant which connects the higher orders susceptibilities to the first order expressions is taken from experiments and is considered to be independent of frequency as in the oscillator model described above.

In our simplified model we obtain as an immediate byproduct Miller's rules to all orders of harmonics generation and can derive the explicit expressions for Miller's constants, which turn out to be independent of frequency.

Comparing eq. (7) and eq. (6) we obtain for the second harmonics:

$$\chi^{(2)}(\omega,\omega) = \Delta_M \chi^{(1)}(2\omega)\left(\chi^{(1)}(\omega)\right)^2 \quad , \tag{12a}$$

with

$$\Delta_M = -\frac{1}{2!}\frac{m}{e^3 N^2}\left[\frac{\partial^3 V(x)}{\partial x^3}\right]_0 \quad ; \tag{12b}$$

where as usual the derivative is computed at the equilibrium in the classical case and as an average on the ground state in the quantum treatment. This is in agreement with the result obtained by Garrett and Robinson [11] and also agrees with the result of Scandolo and Bassani [6].

Our model gives a Miller's rule also for the third harmonic generation. Using eq. (8) and eq. (6) we obtain:

$$\chi^{(3)}(\omega,\omega,\omega) = \Delta_1 \chi^{(1)}(3\omega)\chi^{(1)}(\omega)\chi^{(1)}(\omega)\chi^{(1)}(\omega) +$$

$$+ \Delta_2 \chi^{(1)}(3\omega)\chi^{(1)}(2\omega)\chi^{(1)}(\omega)\chi^{(1)}(\omega)\chi^{(1)}(\omega) \quad , \tag{13a}$$

with:

$$\Delta_1 = -\frac{1}{3!}\frac{m}{e^4 N^3}\left[\frac{\partial^4 V(x)}{\partial x^4}\right]_0 \quad , \tag{13b}$$

and:

$$\Delta_2 = \frac{m^2}{e^6 N^4}\left[\frac{\partial^3 V(x)}{\partial x^3}\right]_0^2 \quad . \tag{13c}$$

The expression for the contribution of the first term in (13b) has been obtained by Scandolo and Rapapa [7] applying the sum rules on a simplified quantum model with a single resonance, provided the ground state average is considered.

Similar expressions can be derived for higher orders harmonic generation susceptibilities, as can be obtained from our equations (9) and (6). The main term, which is responsible for the asymptotic behavior is:

$$\chi^{(n)}(\omega,...,\omega) \approx \Delta_1^{(n)} \chi^{(1)}(n\omega)\left(\chi^{(1)}(\omega)\right)^n \quad , \tag{14a}$$

with:

$$\Delta_1^{(n)} = -\frac{1}{n!}\frac{m}{e^{n+1}N^n}\left[\frac{\partial^{n+1}V(x)}{\partial x^{n+1}}\right]_0 . \tag{14b}$$

Other terms can be obtained by expanding: they can be expressed as products of $\chi^{(1)}(n\omega)\left(\chi^{(1)}(\omega)\right)^n$ times a number of susceptibilities $\chi^{(1)}(i\omega)$ with i<n. The corresponding constants are products of derivatives of the potential of order lower than (n+1) with appropriate dimensional coefficients.

Only the dominant term of Miller's rule (14) is connected with the sum rules because it is the only term responsible for the asymptotic behavior. In the case of inversion symmetry the relevance of (14a) is even greater since all the other terms with odd derivatives in the potential vanish by symmetry.

From equation (14) and equation (11) we see the connection between Miller's constant and the sum rules. That is:

$$\int_0^\infty \omega'^{2n+1}\operatorname{Im}\chi^{(n)}(\omega'...\omega')d\omega' = \frac{\pi}{2}\frac{(-1)^n}{n^2}\frac{e^{2n+2}N^{n+1}}{m^{n+1}}\Delta_1^{(n)} \tag{15}$$

The generality of this result makes it independent of the specific model and allows a connection between the highest momentum of the susceptibilities and Miller's constant, which thus turns out to give a measure of the strength of the nonlinearity.

## V. IONIC AND ELECTRONIC NONLINEAR EFFECTS

The practical application of the oscillator model to physically relevant problems, in the case of polar materials, requires considering both electronic resonances and ionic vibrations.

As a first approximation we can separate the electronic and ionic motions, considering the fact that their resonance frequencies are well separated (in the far infrared for the vibrational modes and in the visible range for the electronic resonances). In this case also the susceptibilities can be separated in two similar contributions, $\chi_e^{(n)}(\omega,\omega..\omega)$ and $\chi_i^{(n)}(\omega,\omega..\omega)$, each one related only to its own parameters (charge, mass, density) and to its own potential. Consequently we obtain two sets of dispersion relations and sum rules of the type (10) and (11), one for the electronic contribution and the other for the ionic contribution, where in the latter case we use the ionic charge Q and the ionic mass M, and the derivatives are evaluated with respect to the ionic coordinate X. The density $N_i$ is also referred to the ions in this case.

We must however consider the fact that such a separation is far from being rigorous, and mixed derivatives of the potential appear already in the second order, as proved by Garrett [11]. We can show that the harmonic generation susceptibilities can be obtained to all orders, including mixed derivatives of the potential energy.

Also in this case K.K. relations can be obtained, considering the analytic properties and the asymptotic behavior of the susceptibilities. They are formally equal to expressions (10), provided we use the expression :

$$\chi^{(n)}(\omega,\omega..\omega) = \chi_e^{(n)}(\omega,\omega..\omega) + \chi_i^{(n)}(\omega,\omega..\omega), \qquad (16)$$

Analogously, sum rules similar to (11) can be obtained for the total susceptibility by considering the asymptotic behaviors which result from the expressions of electronic and ionic susceptibilities. The two contributions have a very similar expression: here we present the $\chi_e^{(n)}(\omega,\omega..\omega)$'s formula:

$$\chi_e^{(n)}(\omega\ldots\omega) = -\sum_{a=0}^{n} \left( \frac{1}{a!(n-a)!} \frac{e^{a+1} Q_i^{n-a} N_e}{m_e^{a+1} M_i^{n-a}} \left[ \frac{\partial^{n+1} U(x_e, X_i)}{\partial x_e^{a+1} \partial X_i^{n-a}} \right]_{(0,0)} \times \right.$$

$$\left. \times \frac{1}{D_e(n\omega) D_e^a(\omega) D_i^{n-a}(\omega)} \right) +$$

$$- \frac{1}{D_e(n\omega)} \left( \sum_{i,k,l,z} \frac{1}{m_e e^{\left(\sum_{y=1}^{l} k_q - 1\right)} Q_i^{\left(\sum_{o=l+1}^{z} k_s\right)} N_e^{\left(\sum_{t=1}^{l} k_r - 1\right)} N_i^{\left(\sum_{r=l+1}^{z} k_t\right)}} \times \right.$$

$$\times \left[ \frac{\partial^{\left(\sum_{w=1}^{z} k_w + 1\right)} U(x_e, X_i)}{\partial x_e^{\left(\sum_{q=1}^{l} k_q + 1\right)} \partial X_i^{\left(\sum_{r=l+1}^{z} k_r\right)}} \right]_{(0,0)} \prod_{j=1}^{l} \frac{\left(\chi_e^{(i_j)}(\omega...\omega)\right)^{k_j}}{k_j!} \cdot \prod_{m=l+1}^{z} \frac{\left(\chi_i^{(i_m)}(\omega...\omega)\right)^{k_m}}{k_m!} \cdot$$
(17a)

with the constraints:

$$\sum_{h=1}^{z} k_h i_h = n \tag{17b}$$

$$\begin{cases} \forall s, s' | 1 \leq s < s' \leq l, \quad \forall w, w' | 1 \leq w < w' \leq z \\ (1 \leq i_s < i_{s'} < n) \wedge (1 \leq i_w < i_{w'} < n) \wedge (0 \leq k_s, k_w \leq k_s + k_w < n). \end{cases} \tag{17c}$$

Note that the third constraint of the (17c) is needed to separate the first term in (17a) from the other ones. The asymptotic behaviour is given by the terms in the first summation in the formula (17a) because they have the slowest decrease at infinity.

Obviously, the sum rules involving all the momenta except the highest do not change because they vanish. The last of the sum rules (11) is modified by the mixed derivatives terms and results to be the following:

$$\int_0^\infty \omega^{2\alpha+1} \operatorname{Im} \chi^{(n)}(\omega,..,\omega) d\omega = \frac{\pi}{2} \frac{1}{n!} \frac{(-1)^{n+1}}{n^2} \frac{e^{n+1} N_e}{m^{n+1}} \left[ \frac{\partial^{n+1} U(x,X)}{\partial x^{n+1}} \right]_{(0,0)} +$$

$$+ \frac{\pi}{2} \frac{1}{n!} \frac{(-1)^{n+!}}{n^2} \frac{Q^{n+1} N_i}{M^{n+1}} \left[ \frac{\partial^{n+1} U(x,X)}{\partial X^{n+1}} \right]_{(0,0)}$$

$$\frac{\pi}{2} \frac{(-1)^{n+1}}{n^2} \left( \sum_{a=0}^{n-1} \left[ \frac{\partial^{n+1} U(x,X)}{\partial x^{n-a} \partial X^{a+1}} \right]_{(0,0)} \frac{e^{n-a} Q^{a+!}}{m^{n-a} M^{a+1}} \times \right.$$

$$\left. \times \left( \frac{N_i}{a!(n-a)!} + \frac{N_e}{(a+1)!(n-a-1)!} \right) \right). \tag{18}$$

The relevance of the mixed terms in the sum rules depends on the material considered. It is to be expected that in soft matter (organic materials) or in ferroelectric materials their influence is considerable.

New Miller's rules can be obtained also in this case considering leading term of (17) which can be expressed in terms of products of $\chi_e^{(1)}(p\omega)$ and $\chi_i^{(1)}(q\omega)$ with $1 \leq p, q < n$ and a factor which can be either $\chi_e^{(1)}(n\omega)$ or $\chi_i^{(1)}(n\omega)$. Also in this case the sum rule (18) can thus be expressed in terms of the appropriate Miller's constants.

## VI. CONCLUSIONS

We may summarize the results of the present work as follows.

Susceptibilities of harmonic generation processes to all orders have been obtained in terms of the anharmonicity in the potential. Every harmonic susceptibility can be expressed in terms of the lower ones, so that by induction we can say it is possible to find an explicit expression of the susceptibilities as sums of products of first order susceptibilities, which is the essence of Miller's rules.

A different number of K.K. relations is obtained for each order of susceptibilitiy, as determined by the asymptotic behavior.

Sum rules on the momenta of the susceptibility have been obtained, the highest momentum only, the $(2n+1)^{th}$ momentum of the imaginary part, being different from zero.

The role of ionic and electronic coupling has been established, and expressions are given in terms of the softness of the material.

While the explicit expressions of the susceptibility are model dependent, the general results obtained concerning K.K. relations, sum rules and Miller's rules have a more general validity, as already proved for the case of second and third harmonic generation. We hope that the above results will be of practical use, as necessary constraints of any model, needed in the experimental development of high harmonics generation.

## ACKNOWLEDGEMENTS

We wish to thank Giuseppe La Rocca for his advice in the course of this work; we are also indebted to Sandro Scandolo, Roberto Cingolani and Marco Bellini for having brought to our attention a number of relevant contributions.